\documentclass[aps,longbibliography,showpacs,twocolumn,superscriptaddress,amsmath,amssymb,verbatim]{revtex4-2}
\usepackage[version=4]{mhchem}

\usepackage{graphicx}
\usepackage{lmodern}
\usepackage{epstopdf}
\usepackage{dcolumn}
\usepackage{bm}
\usepackage{amsmath} 
\usepackage[pdftex,colorlinks=true,citecolor=blue,linkcolor=blue,urlcolor=blue,bookmarks=true]{hyperref}

\usepackage[charter,greekuppercase=italicized]{mathdesign}
\graphicspath{{./}{figures/}}
\usepackage{microtype}

\usepackage{color}
\usepackage{textcomp}
\definecolor{red}{rgb}{1,0,0}

\definecolor{blue}{rgb}{0,0,1}

\definecolor{green}{rgb}{0,1,0}

\newcommand{\tcr}[1]{\textcolor{black}{#1}}
\newcommand{\tcm}[1]{\textcolor{black}{#1}}

\begin{document}
	\preprint{APS}

\title{Spin-liquid-like spin dynamics in the frustrated    antiferromagnet TbBO$_{3}$}

\author{J. Khatua}
\affiliation{Department of Physics, Indian Institute of Technology Madras, Chennai 600036, India}
\author{D. Tay}
\affiliation{Laboratorium für Festkörperphysik, ETH Zürich, CH-8093 Zurich, Switzerland}
\author{T. Shiroka}
\affiliation{Laboratorium für Festkörperphysik, ETH Zürich, CH-8093 Zurich, Switzerland}
\affiliation{\tcr{Center for Neutron and Muon Sciences,} Paul Scherrer Institut, CH-5232 Villigen PSI, Switzerland}
\author{M. Pregelj}
\affiliation{Jo\v{z}ef Stefan Institute, Jamova cesta 39, 1000 Ljubljana, Slovenia}
\affiliation{Faculty of Mathematics and Physics, University of Ljubljana, Jadranska u. 19, 1000 Ljubljana, Slovenia}
\author{K. Kargeti}
\affiliation{Department of Physics, Bennett University, Greater Noida, Uttar Pradesh 201310, India}
\author{S.~K.~Panda}
\affiliation{Department of Physics, Bennett University, Greater Noida, Uttar Pradesh 201310, India}
\author{G. B. G. Stenning}
\affiliation{ISIS Facility, Rutherford Appleton Laboratory, Chilton, Didcot, Oxon OX11 0QX, United Kingdom }
\author{P. Manuel}
\affiliation{ISIS Facility, Rutherford Appleton Laboratory, Chilton, Didcot, Oxon OX11 0QX, United Kingdom }
\author{M. D. Le}
\affiliation{ISIS Facility, Rutherford Appleton Laboratory, Chilton, Didcot, Oxon OX11 0QX, United Kingdom }
\author{D. T. Adroja}
\affiliation{ISIS Facility, Rutherford Appleton Laboratory, Chilton, Didcot, Oxon OX11 0QX, United Kingdom }
\affiliation{Highly Correlated Matter Research Group, Physics Department, University of Johannesburg, Auckland Park 2006, South Africa}
\author{P. Khuntia}
\email[]{pkhuntia@iitm.ac.in}
\affiliation{Department of Physics, Indian Institute of Technology Madras, Chennai 600036, India}
\affiliation{Quantum Centre of Excellence for Diamond and Emergent Materials, Indian Institute of Technology Madras,
Chennai 600036, India}

\date{\today}

\begin{abstract}
The synergistic interplay between spin correlations, spin-orbit
coupling, and competing exchange interactions provides a promising route to realize exotic quantum states
with nontrivial excitations in rare-earth based frustrated magnets. \tcr{Here, \tcm{by using} thermodynamic and local-probe measurements down to 16\,mK, we} demonstrate the exotic magnetism and spin dynamics \tcr{in the} distorted triangular lattice TbBO$_{3}$.  Thermodynamic experiments reveal the presence of dominant antiferromagnetic exchange and subdominant dipolar interactions. Despite 
sizable
antiferromagnetic exchange interact\tcr{ions} between \tcr{the} Tb$^{3+}$ moments, muo\tcr{n-s}pin relaxation experiment does not detect \tcr{any}  signatur\tcr{es} 
of long-range magnetic order \tcr{or} spin-freezing down to 16\,mK, corroborating
\tcm{the} specific heat and \textit{ac} magnetic susceptibility down to 45 mK that suggests a persistent spin dynamics  in this frustrated triangular lattice. 
The scaling of muon relaxation rate as a function of \tcr{the} characteristic energy scale for several spi\tcr{n-l}iquid candidates, including TbBO$_{3}$,  demonstrates that a common underlying mechanism is at play.
The persistent dynamics in this frustrated triangular lattice antiferromagnet is reminiscent of a universal spin-liquid-like spin fluctuations, here attributed to dominant two dimensional (2D) antiferromagnetic short-range spin correlations, confirmed by the presence of a broad magnetic diffuse scattering in the elastic and low-energy inelastic neutron scattering channels at $Q \approx 1.03 $ \AA$^{-1}$ at low temperatures. Our results demonstrate that non-Kramers ion based triangular lattice 
 hosts spin-liquid-like dynamics of local moments arising from the admixture of excited crystal electric field states into the ground state and intertwining of frustration and spin-orbit interaction.
\end{abstract}
\maketitle
Quantum materials with geometrical frustration and competing degrees of
freedom can harbor rich many-body quantum phenomena. They offer a viable platform  to address some of the fundamental questions and \tcm{can act as} key ingredients
for realizing robust quantum-computing technologies~\cite{doi:10.1146/annurev-conmatphys-020911-125138,Balents2010,Jeon2024}. In this vein, quantum spin liquids (QSLs) \tcm{represent}
an elusive state of matter, characterized by the \tcm{lack} of \tcm{a}
symmetry-breaking phase transition down to $T \to 0$, \tcm{by} topological
entanglement entropy, \tcm{a} high quantum entanglement, exotic fractionalized
excitations, \tcm{including} charge-neutral \tcm{spinons,} visons, and
Majorana fermions~\cite{Balents2010,doi:10.1126/science.aay0668,Khuntia2020,PhysRevB.96.235113,KITAEV20062,Balents2010,KHATUA20231,Takagi2019}.

The experimental realization of QSLs with exotic topological  excitations in archetypical frustrated magnets, such as kagome, triangular, and honeycomb lattices 
is highly topical in modern condensed matter~\cite{Savary_2016}. In this context, the triangular lattice (TL) antiferromagnet represents 
a quintessential model for the experimental realization of QSL states. In case of isotropic Heisenberg nearest-neighbor interactions, this lattice features a 120$^{\circ}$ spin-ordered ground state~\cite{PhysRevLett.69.2590}. 
However, the presence of next-neighbor interactions, magnetic anisotropy, or anisotropic exchange interactions can lead to a quantum disordered state in  TL representatives~\cite{PhysRevB.93.144411,PhysRevB.93.140408}. Currently, research efforts have been devoted to $4f$ rare-earth-based frustrated antiferromagnets, where an intricate interplay between strong spin-orbit coupling and crystal electric field (CEF) often leads to a strong exchange anisotropy, thus offering an alternate 
route to realize the QSL state. This route is based on the spin-orbit coupling which, for Kramers ions,
typically hosts
$J_{\rm eff} = 1/2$ as the lowest Kramers doublet state
at low temperature, accompanied by a large magnetic anisotropy due to crystal-field effects~\cite{PhysRevB.92.041105,PhysRevB.94.121111,PhysRevB.98.220409,Bordelon2019,PhysRevX.11.021044,KITAEV20062,PhysRev.79.357,Arh2022}.

\begin{figure}[b] 
	\centering
	\includegraphics[width=0.45\textwidth]{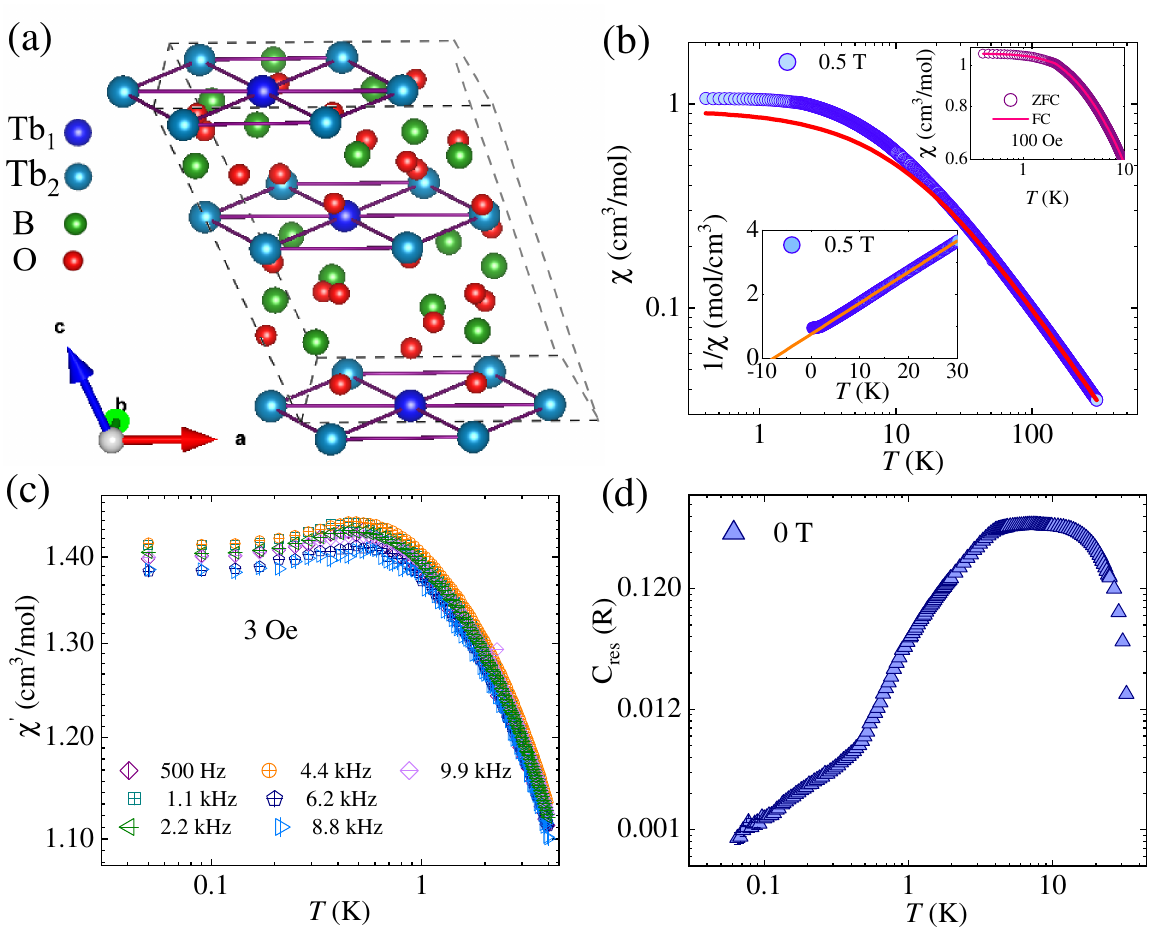}
	\caption{\label{TBOM}(a) Schematic of the distorted triangular lattice formed by Tb$^{3+}$ ions within the $ab$ planes of the monoclinic TbBO$_{3}$ structure. (b) Temperature dependence of  magnetic susceptibility 
		\tcr{collected at} 0.5\,T \tcr{and its} Curie-Weiss fit (solid red line). The top inset shows the temperature dependence of the zero-field-cooled  and field-cooled magnetic susceptibilities in 100 Oe. 
		The bottom inset depicts the Curie-Weiss fit to the low-temperature inverse magnetic susceptibility at 0.5 T. (c) Temperature dependence of \tcr{the} real part of \textit{ac} susceptibility at different frequencies down to 50\,mK. 
		(d) Temperature dependence of \tcr{the} residual specific heat ($C_{\rm res}$) in  zero-field.}
\end{figure}

\begin{figure}[t]
	\centering
	\includegraphics[width=0.5\textwidth]{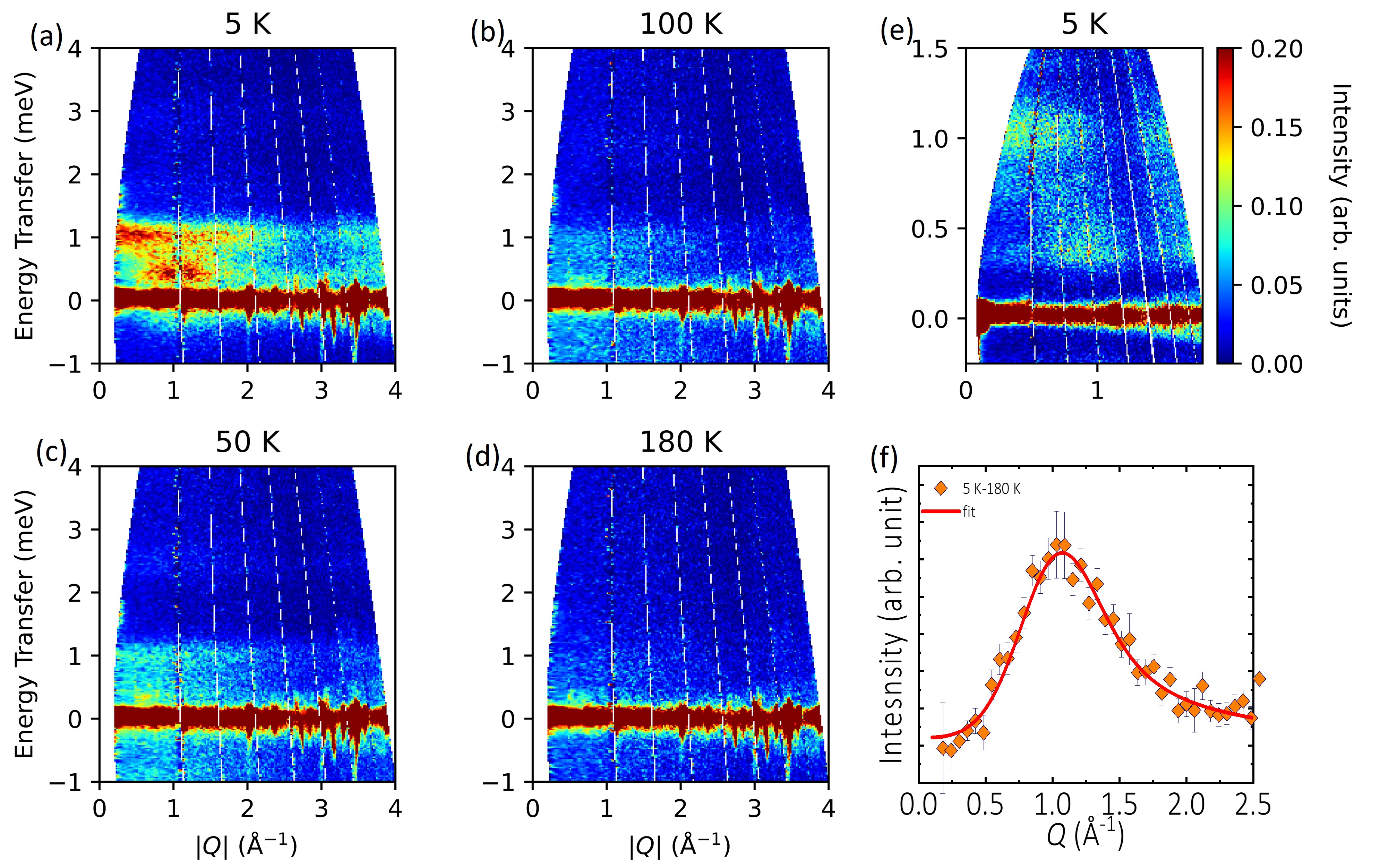}
	\caption{(a-d) Temperature evolution of the inelastic neutron scattering
		spectra recorded on the MARI spectrometer measured with $E_i = 9.2$\,meV over
		the temperature range 5--180\,K. (e) Magnetic diffuse scattering at
		5\,K $\approx$ $ \textcolor{black}{|\theta_\mathrm{CW}^{\rm L}|}$ in the
		low-energy inelastic neutron scattering measured with  $E_i $= 2 meV. (f) Energy integrated 
		($\Delta E = 0.25$--0.5)\,meV, momentum ($Q$) dependent intensity at 5\,K for 4 meV after subtracting the background contribution  at  $T$ = 180 K  and the solid
		line represents a fit corresponding to the Warren function.}
	\label{INS_d} 
\end{figure}

Frustrated magnets containing non-Kramers 4$f$-ions with an integer total angular momentum $J$  can also
host 
various 
many-body phenomena, such as quadrupolar order~\cite{doi:10.7566/JPSJ.83.034709}, coexistence of quadrupolar orders
and intertwined multipolar orders~\cite{PhysRevB.98.045119},
spin–orbital liquid state~\cite{Tang2023}, perturbation induced quantum criticality, dispersion of magnetic exciton modes~\cite{PhysRevB.109.115110}, and Ising nematic order~\cite{doi:10.1073/pnas.2119942119,Vinograd2022}. While non-Kramers systems ideally manifest a nonmagnetic singlet ground state, a key query is whether they can stabilize spin-liquid-like state similar to the $J_{\rm eff}=1/2$ Kramers magnets. Recent theoretical studies suggest that non-Kramers ion based magnet can indeed harbor induced quantum magnetism, where quantum superpositions between the singlet ground state and low-lying excited states generate effective local magnetic moments, even in the absence of disorder~\cite{PhysRevB.109.115110,Sibille2017,PhysRevResearch.6.023267}. Nevertheless, the role of the singlet–excited-state gap, anisotropic magnetic interactions, and lattice distortions on the underlying grounds state, remains unclear. To gain deeper insight into such quantum phenomena in non-Kramers  magnets, it is therefore essential to investigate  promising non-Kramers ion based frustrated magnets. In this context, disorder free frustrated magnets on lanthanide compound with non-Kramers ion renders an appealing venue to explore induced quantum magnetism.\\ \\
Here, we investigated the ground state of a distorted triangular lattice antiferromagnet TbBO$_{3}$ by  complementary  muon spin relaxation ({\textmu}SR), nuclear magnetic resonance (NMR), inelastic neutron scattering (INS), neutron powder diffraction (NPD), and thermodynamic experiments. Our experiments along with density functional theory + Hubbard U (DFT + U) calculations suggest the presence of \tcr{a} significant spin-orbit driven magnetic anisotropy and a frustrated triangular lattice. The \textit{ac} magnetic susceptibility, specific-heat, and NPD  \tcr{data} down to 50\,mK rule out \tcr{the presence of} spin freezing and magnetic phase transitions. Inelastic neutron scattering experiments reveal a tiny spin gap of $\sim 3.5$~K in the low-energy excitation spectra~\cite{PhysRevLett.84.2957,PhysRevB.110.L020402}.
Despite the 
\tcr{sizable} antiferromagnetic interaction between Tb$^{3+}$  moments, {\textmu}SR experiments detect no signatures of long-range magnetic order or spin-freezing down to 16\,mK. The $\mu$SR relaxation rate remains nearly constant below 1 K, a behavior commonly observed in spin-liquid candidates and indicative of a dynamic ground state \cite{Clark2019}. The low-energy inelastic excitations \tcm{at} 0.3 and 0.5\,meV conducted on MARI spectrometer at ISIS, UK \tcm{show} a broad maximum at $Q\approx 1.03$ \AA$^{-1}$ that is enhanced at low temperatures $T \sim \textcolor{black}{|\theta_\mathrm{CW}^{\rm L}|}$ (see below), consistent with magnetic diffuse elastic scattering at 10 K taken on WISH time of flight neutron diffractometer (see supplementary material
	 [SM]~\cite{sm}).
The broad diffuse scattering both in elastic and inelastic channels
\tcm{corroborates} a spin-liquid-like state with a  dominant antiferromagnetic  2D short-range spin correlations with correlation length of $\sim$10\,\AA. Our comprehensive results uncover short-range spin correlations and persistent spin dynamics well below the interaction energy scale, supporting a spin-liquid-like spin dynamics in this frustrated magnet.\\ \\
The Rietveld refinement of the powder XRD data \tcr{(see Fig.~S1 in SM~\cite{sm})} reve\tcr{als} 
that TbBO$_{3}$ crystallizes in the monoclinic $C2/c$ space group. The obtained lattice parameter and atomic coordinates (see \cite{sm}) 
 confirm the absence of anti-site disorder \tcr{among the}  
constitu\tcr{ent} atoms. Figure~\ref{TBOM}(a) depicts the monoclinic crystal structure of TbBO$_{3}$~\cite{Lin2004,MUKHERJEE2018173,Lin2024}, 
\tcr{where the Tb$^{3+}$ ions occupy two distinct Wyckoff sites: $4c$ (Tb$_{1}$) and $8f$ (Tb$_{2}$).}  
The Tb$^{3+}$ ions at the $4c$ sites occupy at the hexagon centers of the distorted
honeycomb lattice by the Tb$^{3+}$ ions at the $8f$ sites and form a 
distorted triangular lattice, with Tb--Tb--Tb bond angles ranging from
$59.15^{\circ}$ to $60.26^{\circ}$ ( Fig.~\ref{TBOM}(a))~\cite{Lin2004}.
The NPD experiments on WISH reveal the existence of significant peak broadening not observed in our XRD (not shown here). The scattering lengths of B and O are much smaller for X-rays than for neutrons, and this disorder is therefore likely associated with the borate framework~\cite{mukherjee_2018}, which cannot be well reproduced by standard strain modeling implemented in Fullprof. Further exploration of this structural broadening would require single-crystal diffraction, which is beyond the scope of this study. Nonetheless, NPD reveals the absence of anti-site disorder,  rules out magnetic ordering down to 50~mK, and detects magnetic diffuse scattering at low temperatures$\sim$ $|\theta_\mathrm{CW}^{\rm L}|$  (see \cite{sm}). \\
\begin{figure*}[htbp]
	\centering
	\includegraphics[width=\textwidth]{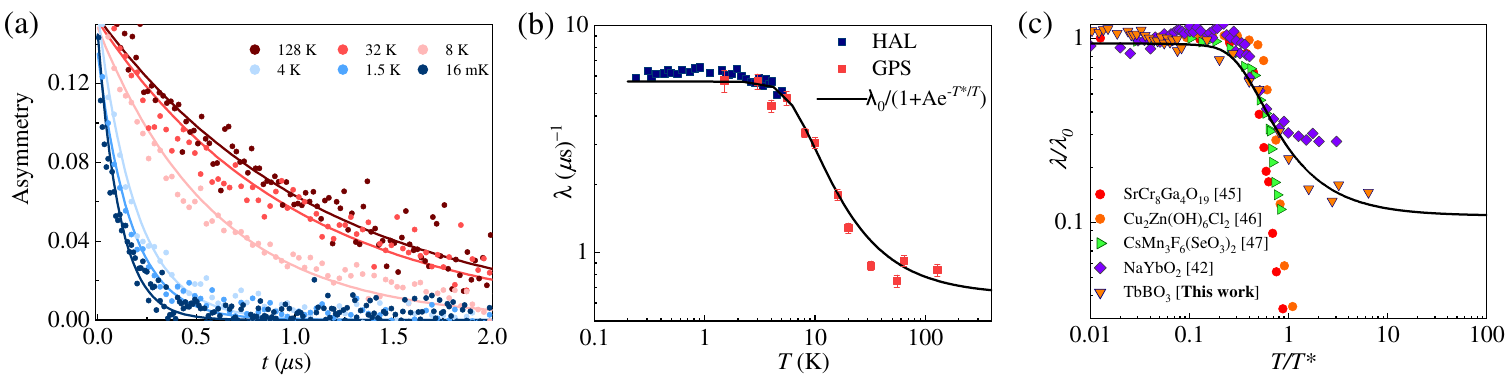}
	\caption{\tcr{(a) Time evolution of the \tcr{{\textmu}SR} asymmetry
			at representative  temperatures. Solid lines represent fits to an
			exponential function, as described in the text. (b) Temperature dependence
			of the muon-spin relaxation rate $\lambda$. To cover a broad temperature
			range, the datasets had to be collected at two different {\textmu}SR
			spectrometers: HAL and GPS. (c) Temperature-dependent scaling of $\lambda$
			for several potential spin-liquid materials. The solid line in (b) and (c)
			corresponds to the phenomenological model
			$\lambda = \lambda_{0}/[1+ A \exp(-T^{*}/T)]$, which captures the
			thermally activated behavior of magnetic moments, with $\lambda_{0}$ and $A$
			two constants and $T^{*} = 20 \pm 6$\,K a characteristic energy scale.}}
	\label{fig:muSR}
\end{figure*}
As shown in Fig.~\ref{TBOM}(b), the absence of any anomalies in the temperature dependence of $dc$ magnetic susceptibility $\chi(T)$ down to 0.4 K rules out 
\tcr{a} long-range magnetic order (LRO) \tcr{of} Tb$^{3+}$ \tcr{moments.}
\tcr{Further, the overlapping} 
zero-field cooled and field-cooled $\chi(T)$ data \tcr{imply} that Tb$^{3+}$ spins \tcr{do not freeze,} at least down to 0.4\,K (see inset in Fig.~\ref{TBOM}(b)). \tcr{The lack of spin freezing is further supported by low-temperature} \textit{ac} magnetic susceptibility measurements performed at various  frequencies (see Fig.~\ref{TBOM}(c)). 
The real part of ac susceptibility $\chi^{'}(T)$ exhibits a broad hump 
around 0.7\,K, \tcr{independent of frequency,} similar to other
sp\tcr{in-li}quid candidates~\cite{Sarte2021}.
To estimate the 
effective magnetic moment $\mu_{\rm eff}$ and \tcr{the} Curie-Weiss (CW) temperature $\theta_{\rm CW}$, the inverse magnetic susceptibility data 
1/$\chi(T)$ were fitted by the CW law $\chi = C/(T-\theta_{\rm CW}$).   
\tcr{Here,} \tcr{the} Curie constant $C$ is related to the effective moment 
by $\mu_{\rm eff} =\sqrt{8C} \mu_{\rm B}$, while $\theta_{\rm CW}$ represents the energy scale of \tcr{the} magnetic exchange interaction between Tb$^{3+}$ moments. The CW fit to the high-temperature ($T> 50$\,K) 1/$\chi$ data (red line in Fig.~\ref{TBOM}(b)) yields an effective moment of 9.62\,$\mu_{\rm B}$, \tcr{close} to the theoretical value of 9.72\,$\mu_{\rm B}$ of \tcr{the} free Tb$^{3+}$ ions ($^7$F$_{6}$: $S = 3$, $L = 3$). \tcr{The CW} temperature $\textcolor{black}{|\theta_\mathrm{CW}^{\rm H}|} = -12$\,K suggests the presence of 
low-energy excited CEF levels, consistent with our INS experiment that yield first excited level at $\sim$1 meV (11.6 K), which is in agreement with a maximum  at $T\sim$ 6 K in residual specific heat,  $C_{\rm res}$ (see Fig.~\ref{INS_d}(a)-(e) and SM). To obtain a rough estimate 
of the dominant magnetic interactions between Tb$^{3+}$ moments, the 
1/$\chi$ data below 25\,K (bottom inset of Fig.~\ref{TBOM}(b)) were fitted by \tcr{a} CW law, \tcr{giving} 
$\mu_{\rm eff} = 9.06$\,$\mu_{\rm B}$, and $\textcolor{black}{|\theta_\mathrm{CW}^{\rm L}|}$  = $-7.6$ K, suggesting dominant antiferromagnetic interaction between Tb$^{3+}$ moments. \tcr{The} effective moment derived 
from a CW fit to the low-temperature $\chi(T)$ data is  
\tcr{similar} to that obtained \tcr{at} high-temperature. This suggests the role of spin-orbit driven anisotropy in the exchange interaction in inducing frustration in the present system~\cite{sm}. In addition, a dipolar interaction $J_d=  0.94 ~\mathrm{K}$ between Tb$^{3+}$ moments separated by an average distance of 3.78~Å  is active in this triangular lattice antiferromagnet (see \cite{sm}). The strong spin-orbit coupling, CEF, distortion in the spin-lattice, and easy-axis anisotropy induce magnetic frustration in TbBO$_{3}$ ~\cite{Lin2024}.
\\ In TbBO$_{3}$, the character of a local moment at extremely low temperatures is expected to arise from the quantum superposition of the singlet ground state and the excited state. Depending on the strength of the magnetic exchange and CEF splitting between the two singlets, the system will exhibit either induced moment magnetic ordering or a cooperative paramagnetic ground state (i.e., no magnetic ordering) \cite{PhysRevB.109.115110,Goremychkin2008}. A criterion, defined by a dimensionless control parameter, $\xi
	 = m^{2}J_{\rm ex}/2\Delta$, with $m$ as the effective magnetic moment, $J_{\rm ex}$ as the exchange interaction, and $\Delta$ as the local CEF gap, was proposed~\cite{PhysRevB.109.115110}. A quantum critical point is predicted for $\xi$ = 1. For $\xi < 1$, no magnetic ordering occurs, while if $\xi > 1$, then the exchange interaction is strong enough to admix the excited state into the ground state, thereby generating a quasi-doublet with finite magnetic character~\cite{PhysRevB.109.115110}. This should result in magnetic ordering at some finite temperature, where the admixed state would realize a pseudospin-1/2 degree of freedom and manifest as reduced effective moments at low temperatures, for example, in TbInO$_{3}$ \cite{Clark2019}. However, with our values of the deduced ground state CEF matrix element $m$ = 2.162, exchange interactions $J_{\rm ex}$ = 2.2 K, and CEF gap $\Delta$ = 11.6 K, leading to $|\xi| < 1$, the frustrated magnet TbBO$_{3}$ should never order.  Furthermore, our measurements reveal that the present system retains unusually large effective moments even at low temperatures. This observation suggests that extended superexchange and thermal excitation into the low-lying excited state play a dominant role in the large moment $\sim 6$ $\mu_{\rm B}$ at a low temperature of $\sim 5 $ K, rather than a scenario dominated only by quantum superposition between the singlet and excited states~\cite{PhysRevLett.116.097205}. \\
\tcr{To obtain further insight} into \tcr{the} magnetic correlations 
and low-energy excitations, we performed specific-heat $C_{p}$ measurements down to 65\,mK in zero-field. The absence of any signature of $\lambda$-type anomaly in $C_{p}$ suggests
that the Tb$^{3+}$ moments do not undergo a symmetry breaking LRO, at least down to
65\,mK (Fig.~\ref{TBOM}(d) and Fig.~S4 in~\cite{sm}).
Following the subtraction of the lattice  and nuclear contributions (see Fig.~S4),
the residual specific heat, i.e., $C_{\rm res}(T) = C_{p}(T)-C_{\rm lat.}(T) -C_{\rm nuc.}$,
is obtained and \tcr{it is} shown in Fig.~\ref{TBOM}(d).
In zero-field, \tcr{$C_{\rm res}$ shows} a broad maximum over a wide
temperature range, similar to other rare-earth-based magnets \cite{PhysRevX.9.031005}.
Nonetheless, in case of TbBO$_{3}$, \tcr{this maximum (ranging from
4\,K to 25\,K)} is attributed to an overlap of the Schottky anomaly
(owing to the thermal population of the excited CEF levels) and \tcr{the}
short\--ran\-ge spin correlations~\cite{PhysRevX.9.031005}, as supported by our CEF analysis (see \cite{sm}). In addition, our INS experiment detect the first CEF level at 1 meV, which will give rise to a peak in $C_{\rm res}$ at 1 meV/2 = 0.5 meV $\sim$ 5.8 K.
\tcr{Indeed, upon} lowering the temperature, $C_{\rm res}$ starts
\tcr{to increase,} followed by a broad maximum around 5\,K (see Fig.~\ref{TBOM}(d)) 
reflecting the growth of short-range magnetic correlations, in line with that found in many low-dimensional frustrated magnets,
where magnetic correlations can persist even at high temperatures~\cite{PhysRevX.9.031005,l93vf576}. The thermodynamic evidence of short-range magnetic correlations aligns with the observed magnetic diffuse scattering in neutron experiments at low temperatures $T\sim \textcolor{black}{|\theta_\mathrm{CW}^{\rm L}|}$ (see below).
In addition, it reflects that non-Kramers ions are promising contenders for hosting correlated magnetism, consistent with recent theoretical proposals~\cite{PhysRevB.109.115110}.\\

\tcr{To} ascertain the absence of LRO and spin-freezing, \tcr{and to}
unveil \tcr{unambiguously} the spin dynamics in the ground state, we
performed zero-field (ZF) {\textmu}SR measurements on TbBO$_{3}$.
As illustrated in Fig.~\ref{fig:muSR}(a), a rapid relaxation of muon-spin polarization occurs at \tcr{early times} ($t < 2$\,{\textmu}s), \tcr{more
pronounced at lower temperatures.} However, the ZF-{\textmu}SR asymmetry
show\tcr{s} no signature\tcr{s} of oscillation, nor a 1/3 ``tail'' at
long times, thus indicating the absence of internal static magnetic fields at
the muon stopping site~\cite{Clark2019}. \tcr{To follow the evolution of spin dynamics with temperature,} the
depolarization function $A(t) = A_{0} \rm exp (-\lambda t)$
was used to describe the ZF-{\textmu}SR spectra \tcr{over} the entire
temperature \tcr{range,} 0.016--128\,K. Here, $A_0$ is the initial
asymmetry, while $\lambda$ represents the muon-spin relaxation rate.
\tcr{Figure~\ref{fig:muSR}(b) shows} the temperature dependence of
$\lambda$. As the temperatures decreases, $\lambda$ exhibits a rising
trend, until it reaches a plateau at approximately 5\,K, 
\tcr{in agreement} with the
\tcr{broad maximum observed in the} specific heat, $C_{\rm res}$. 
This indicates a slowing down \tcr{of the} spin dynamics, attributed
to the development of short-range spin correlations~\cite{PhysRevB.100.144432}. The saturation of $\lambda$ \tcr{below 1\,K} implies \tcr{a} persistent spin dynamics, a
characteristic feature of the spin-liquid like ground state~\cite{PhysRevB.98.014431,l93vf576,Sibille2017,PhysRevB.100.144432}.

The decrease of $\lambda$ with temperature in a narrow temperature
range (\tcr{6\,K $\leq T \leq$ 30\,K, for TbBO$_{3}$),} is also observed
in \tcr{several other} spin-liquid candidate materials, including SrCr$_{8}$Ga$_{4}$O$_{19}$ \cite{PhysRevB.58.12049,PhysRevLett.73.3306}, Cu$_3$Zn(OH)$_6$Cl$_2$~\cite{PhysRevLett.109.037208}, CsMn$_{3}$F$_{6}$(SeO$_{3}$)$_{2}$ \cite{PhysRevB.105.094439}, and
NaYbO$_2$~\cite{PhysRevB.100.144432}.
To shed \tcr{light on} the low-temperature spin dynamics of \tcr{these}
promising \tcr{spin-liquid} materials and to establish a \tcr{common  framework across
them, the} temperature dependence of $\lambda$ was fitted using
a model of thermally activated behavior (see Fig.~\ref{fig:muSR}(b)) of electronic spins~\cite{PhysRevB.75.094404}, where the characteristic energy scale $T^{*}$ can
be interpreted as a measure of the strength of thermally activated gap related to the CEF gap observed in INS experiments as discussed in the INS section. This suggests that $\mu$SR relaxation mechanism is governed  by Orbach process \cite{le2011muon} with a CEF gap of 20 K.  It contrasts with many 3$d$ systems, where activation typically indicates a thermal gap between low-temperature correlated state and high-temperature paramagnetic state ~\cite{PhysRevB.75.094404}.\\
The normalized {\textmu}SR relaxation rate  for each of the above
\tcr{spin liquid} materials 
is plotted in Fig.~\ref{fig:muSR}(c). \tcr{Here,} the relaxation is
normalized to the respective $\lambda_{0}(T = 0 \ \ \text{K})$ value, while the temperature is scaled
to the characteristic $T^{*}$ value, at which the {\textmu}SR relaxation
begins to increase on cooling.
\begin{figure}[b]
	\centering
	\includegraphics[width=0.48\textwidth]{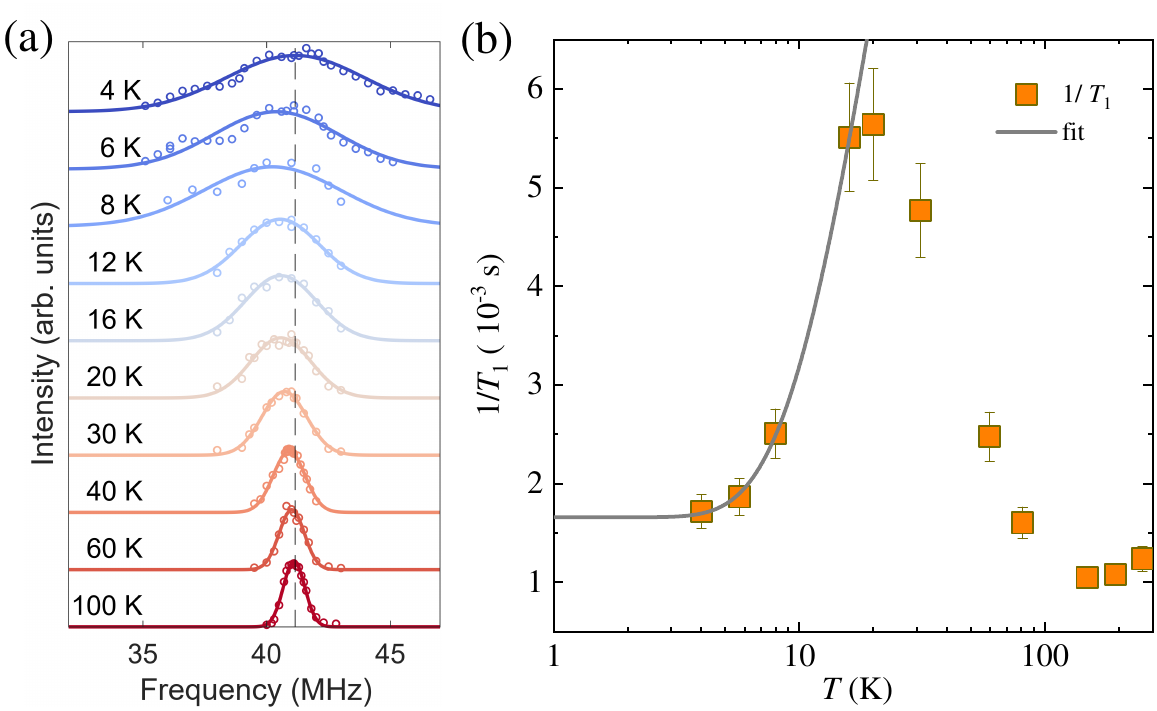}
	\caption{(a) Frequency-swept $^{11}$B NMR spectra measured in a
		\tcr{magnetic field of 3\,T} at various temperatures. Solid lines
		represent the calculated spectra, with the full width at half maximum
		as a fit parameter. The reference frequency is shown by the dashed
		vertical line. (b) $^{11}$B NMR spin-lattice relaxation rate
		($T_{\rm 1}^{-1}$) vs.\	temperature in TbBO$_3$. The solid line is a fit corresponds to a thermally activated behavior below 20 K as discussed in the text.}
	\label{fig:NMR_line}
\end{figure} 
Interestingly, we observe that, even though the normalized {\textmu}SR
relaxation rates
exhibit different temperature dependences above $T^{*}$, \tcr{below it}
all the spin liquid candidates exhibit the same temperature dependence,
\tcr{in this case corresponding to the same fit curve.}
\tcr{This is a strong evidence of} a generic underlying mechanism  that is short-range correlation among fluctuating moments being
at play at low temperatures in all these promising frustrated magnets~\cite{PhysRevLett.116.107203}.  Specifically, although some of the compounds shown in \tcm{Fig.~\ref{fig:muSR}(c)}  have $S = 1/2$, where quantum fluctuations play a prominent role, the present compound, TbBO$_{3}$, hosts large local moments. Despite this, TbBO$_{3}$  exhibits pronounced spin dynamics, suggesting a possible scenario of spin-liquid-like dynamics sustained by frustrated exchange interactions  between the local moments \cite{l93vf576,PhysRevB.108.064432,Sibille2017}. \\ 

To gain microscopic insight into the static susceptibility and
spin dynamics in
the ground state, we performed $^{11}$B
($I = 3/2$, $\tcr{\gamma} = 13.655$\,MHz/T) NMR measurements on the polycrystalline samples of TbBO$_{3}$.
Figure~\ref{fig:NMR_line}(a) depicts the frequency swept $^{11}$B NMR
spectra at several temperatures in \tcr{a fixed} magnetic field $\mu_{0}H = 3$\,T.
The absence of any quadrupole satellite transitions in the measured
frequency range
suggests that the $^{11}$B nucleus undergoes only weak quadrupole
interactions. $^{159}$Tb ($I=3/2$) nucleus, on the other hand, has a sizable quadrupole moment to detect NQR signals at low temperatures; however, carrying out $^{159}$Tb NMR/NQR at dilution refrigerator temperatures is beyond the scope of the current manuscript. In addition, despite the presence of two distinct boron
crystallographic sites in TbBO$_{3}$, the occurrence of a single 
broad peak in the $^{11}$B NMR spectra indicates equivalent environments
for the B nuclei. The broad NMR spectra are possibly related to unavoidable disorder in the borate framework in a polycrystalline sample~\cite{mukherjee_2018}.
The NMR
linewidth broadens significantly, from $\sim 1$\,MHz at 100\,K to
$\sim 8$\,MHz at 4\,K (Fig.~S7)~\cite{sm}. The broadening of the NMR linewidth at
low temperatures is consistent with the enhancement of magnetic susceptibility (Fig.~\ref{TBOM}(b)).
The absence of a rectangular NMR lineshape or of peak splitting rules out
 long-range magnetic order in TbBO$_{3}$, at least down
to 4\,K~\cite{doi:10.1143/JPSJ.55.1751,PhysRevB.102.045149}.
Further, the absence of any significant shifts of the boron frequency 
from its reference value indicates that the hyperfine interaction 
between the boron nucleus and the magnetic Tb$^{3+}$ ions is of dipolar origin. The spin-lattice relaxation rate $T^{-1}_{1}$ probes the low-energy
spin excitations \tcr{reflecting the electron-spin} fluctuations at
the B nucleus site.
The temperature dependence of $T_{1}^{-1}$, derived from the recovery
of the longitudinal nuclear magnetization (see Fig.~\ref{fig:NMR_line}(b)),
reveals a sharp peak around 20\,K. This peak is close to the first
excited CEF gap of $\sim 12$\,K directly observed in our INS study (Fig.~\ref{INS_d}(a)-(e)) and is consistent also with the
{\textmu}SR characteristic energy scale $T^{*} = 20 \pm 6$\,K
(Fig.~\ref{fig:muSR}) representing the CEF gap, which implies that $\mu$SR relaxation is driven by Orbach process.
Below 20\,K, there is a clear decrease of $T_{1}^{-1}$, potentially
associated with the slowing down of spin dynamics due \textcolor{black}{to field-induced admixing of excited CEF states into the singlet ground state, leading to an effective Zeeman response, without any signatures of long-range magnetic order~\cite{PhysRevB.102.045149}.} The NMR relaxation rate below 20 K is reproduced with a thermally activated behavior, $
1/T_{1} = 1/T_1^{0} + A\, \exp\!\left(-\frac{\Delta_{\mathrm{CEF}}}{k_{\mathrm{B}} T}\right)$
 ( Fig.~\ref{fig:NMR_line}(b)), with a residual relaxation rate 1/$T_{1}^{0} = 1.66\times 10^{-3}$  s$^{-1}$ and $\Delta_{CEF} = 24.5$ K~\cite{PhysRevLett.104.057202}. It may be noted that the value of the CEF gap by {\textmu}SR and NMR  is slightly higher than that obtained from INS, given the fact that the gap obtained from the {\textmu}SR and NMR relaxation rates is an indirect method, which is likely to be an overestimate. \\ \\
The INS experiments \tcm{were decisive} to understand the nature of low-energy excitations and spin correlations. At high temperatures, we find no pronounced low-energy magnetic signal.
\tcm{However, as shown in Fig.~\ref{INS_d}, this} becomes apparent below
$\sim 100$\,K. At base temperature (5\,K), comparable with the characteristic
exchange energy scale $ \textcolor{black}{|\theta_\mathrm{CW}^{\rm L}|}$, the emergence of two distinct
\tcm{dispersionless} excitation bands at $\sim 0.3$ and $\sim 1$\,meV
(Fig.~\ref{INS_d}(a)), 
\tcm{implies that} most  \tcm{these} likely reflect \tcm{the} local magnetic properties.
The 1-meV excitation band is suppressed at \tcm{a} higher scattering wave
vector $Q$, \tcm{thus following} the Tb magnetic form factor. \tcm{Further,}
its energy matches the first excited Tb CEF level, \tcm{as resulting from} our calculations (see \cite{sm}).
Hence, we ascribe it to the first excited Tb CEF level.
On the other hand, the 0.3-meV band exhibits a broad maximum at Q\,$\approx$\,1.03\,\AA$^{-1}$ at 5 K (Fig.~\ref{INS_d}(f)), implying dominant antiferromagnetic spin correlations.
Indeed, \tcm{the} modeling of the broad maximum in the magnetic diffuse in the elastic and low-energy inelastic channels
with the Warren function~\cite{Clark2019},
yields a correlation length of \tcm{only} $\sim 10$\,\AA , which is consistent with magnetic diffuse elastic scattering results at 10\,K taken on WISH time-of-flight neutron diffractometer at ISIS, UK (see \cite{sm}).
\tcm{This is} in line with \tcm{the} 2D short-range antiferromagnetic spin correlations 
\tcm{expected in } frustrated magnets \cite{Sibille2017}  \tcm{and, thus, it} confirms
our thermodynamic, NMR, and $\mu$SR results. Furthermore, INS reveal a gapped excitation spectra with a tiny spin gap $\sim 3.5$~K (Fig.~\ref{INS_d}), possibly suggesting topological spin correlations that are expected in  spin liquid candidates ~\cite{Sibille2017,PhysRevB.110.L020402,Niggemann_2020,PhysRevLett.93.167204}.
  \\ \\
In summary, we investigated \tcr{the} spin-orbit-coupled frustrated distorted triangular lattice antiferromagnet TbBO$_{3}$ using thermodynamic
\tcm{and} local-probe techniques. \tcr{The} thermodynamic measurements indicate that the
antiferromagnetically coupled Tb$^{3+}$ moments do not exhibit any
LRO or spin freezing down to 45\,mK, corroborated  by NPD experiments.
The INS experiments and CEF calculations reproduce the ther\-mo\-dy\-na\-mic results reasonably well
and point to a small CEF gap between the ground and excited states.
DFT calculations suggest the presence of a significant spin-orbit
driven magnetic anisotropy. NMR experiments reveal the absence of magnetic ordering and detect a CEF gap, consistent with  {\textmu}SR and INS results. The {\textmu}SR experiments
rule out 
\tcr{a spin freezing} or a magnetic phase transition, but \tcr{confirm} the
persistence of spin dynamics down to 16\,mK. The emergence of
a broad magnetic diffuse contribution in the elastic and low-energy inelastic scattering at $Q = 1.03$\,\AA$^{-1}$ that is enhanced at low-temperature $T\sim$ $\textcolor{black}{|\theta_\mathrm{CW}^{\rm L}|}$  \tcm{implies}  the presence of dominant
antiferromagnetic 2D short range  spin correlations and a spin-liquid-like state, 
\tcm{in agreement with the} {\textmu}SR  and thermodynamic results. The low-energy INS experiments detect a tiny spin gap of ~3.5~K, reflecting topological spin correlations in this frustrated magnet. Our results suggest a possible scenario of  spin-liquid-like spin fluctuations with a gapped excitation spectrum, stabilized by spin–orbit-coupling,  anisotropy, and frustration in TbBO$_{3}$. Future experiments on the  single-crystals  combined with the realistic Hamiltonian modeling that incorporates both dipolar and exchange interactions may add  further credence to this scenario.
\vspace*{0.1 cm}\\

\textit{} 
P. K. acknowledges the
funding by the Anusandhan National Research Foundation
(ANRF), Department of Science and Technology,
India through Research Grants. J. K. acknowledges U. Jena and M. Barik for their support in synthesizing one batch of TbBO3 sample and in zero field specific heat experiment. M.\ P.\
acknowledges funding from the Slovenian Research and Innovation Agency
(ARIS) (Project No.\ J2-2513 and Program No.\ P1-0125). 
 \tcr{We acknowledge the
allocation of beam time at the Swiss muon source (\tcr{HAL} and
GPS {\textmu}SR spectrometers). S.K.P. acknowledges funding support from a SERB Core Research grant (Grant No.  CRG/2023/003063). D.T.A. thanks EPSRC UK (Grant Ref: EP/W00562X/1) for the funding.}

 Data availability: The neutron scattering data that support the findings of the present work are openly available \cite{khuntia25}. All other data are available from the corresponding author upon reasonable request.
 
\bibliography{TbBO3_main}

@article{PhysRevLett.116.097205,
  title = {Origin of the Spin-Orbital Liquid State in a Nearly $\text{J}=0$ Iridate \ce{Ba3ZnIr2O9}},
  author = {Nag, Abhishek and Middey, S. and Bhowal, Sayantika and Panda, S. K. and Mathieu, Roland and Orain, J. C. and Bert, F. and Mendels, P. and Freeman, P. G. and Mansson, M. and Ronnow, H. M. and Telling, M. and Biswas, P. K. and Sheptyakov, D. and Kaushik, S. D. and Siruguri, Vasudeva and Meneghini, Carlo and Sarma, D. D. and Dasgupta, Indra and Ray, Sugata},
  journal = {Phys. Rev. Lett.},
  volume = {116},
  issue = {9},
  pages = {097205},
  numpages = {5},
  year = {2016},
  month = {Mar},
  publisher = {American Physical Society},
  doi = {10.1103/PhysRevLett.116.097205},
}

@article{PhysRevB.98.014431,
  title = {Origin of magnetic moments and presence of spin-orbit singlets in \ce{Ba2YIrO6}},
  author = {Nag, Abhishek and Bhowal, Sayantika and Chakraborty, Atasi and Sala, M. M. and Efimenko, A. and Bert, F. and Biswas, P. K. and Hillier, A. D. and Itoh, M. and Kaushik, S. D. and Siruguri, V. and Meneghini, C. and Dasgupta, I. and Ray, Sugata},
  journal = {Phys. Rev. B},
  volume = {98},
  issue = {1},
  pages = {014431},
  numpages = {7},
  year = {2018},
  month = {Jul},
  publisher = {American Physical Society},
  doi = {10.1103/PhysRevB.98.014431},
}

@article{PhysRevB.93.140408,
  title = {Spin liquid state in the disordered triangular lattice \ce{Sc2Ga2CuO7} revealed by \text{NMR}},
  author = {Khuntia, P. and Kumar, R. and Mahajan, A. V. and Baenitz, M. and Furukawa, Y.},
  journal = {Phys. Rev. B},
  volume = {93},
  issue = {14},
  pages = {140408},
  numpages = {6},
  year = {2016},
  month = {Apr},
  publisher = {American Physical Society},
  doi = {10.1103/PhysRevB.93.140408},
}

@Article{Arh2022,
author={Arh, T.
and Sana, B.
and Pregelj, M.
and Khuntia, P.
and Jagli{\v{c}}i{\'{c}}, Z.
and Le, M. D.
and Biswas, P. K.
and Manuel, P.
and Mangin-Thro, L.
and Ozarowski, A.
and Zorko, A.},
title={The Ising triangular-lattice antiferromagnet neodymium heptatantalate as a quantum spin liquid candidate},
journal={Nat. Mater.},
year={2022},
month={Apr},
day={01},
volume={21},
number={4},
pages={416-422},
issn={1476-4660},
doi={10.1038/s41563-021-01169-y},
}

@article{l93vf576,
  title = {Spin Dynamics in the Dirac U(1) Spin Liquid \text{Yb}\text{Zn$_{2}$}\text{Ga}\text{O$_{5}$} },
  author = {Wu, Hank C. H. and Pratt, Francis L. and Huddart, Benjamin M. and Chatterjee, Dipranjan and Goddard, Paul A. and Singleton, John and Prabhakaran, D. and Blundell, Stephen J.},
  journal = {Phys. Rev. Lett.},
  volume = {135},
  issue = {4},
  pages = {046704},
  numpages = {8},
  year = {2025},
  month = {Jul},
  publisher = {American Physical Society},
  doi = {10.1103/l93v-f576},
}

@Article{Sibille2017,
author={Sibille, Romain
and Lhotel, Elsa
and Ciomaga Hatnean, Monica
and Nilsen, G{\o}ran J.
and Ehlers, Georg
and Cervellino, Antonio
and Ressouche, Eric
and Frontzek, Matthias
and Zaharko, Oksana
and Pomjakushin, Vladimir
and Stuhr, Uwe
and Walker, Helen C.
and Adroja, Devashibhai T.
and Luetkens, Hubertus
and Baines, Chris
and Amato, Alex
and Balakrishnan, Geetha
and Fennell, Tom
and Kenzelmann, Michel},
title={Coulomb spin liquid in anion-disordered pyrochlore \text{Tb$_{2}$\text{Hf$_{2}$}\text{O$_{7}$}}},
journal={Nat. Commun.},
year={2017},
month={Oct},
day={12},
volume={8},
number={1},
pages={892},
issn={2041-1723},
doi={10.1038/s41467-017-00905-w},
url={https://doi.org/10.1038/s41467-017-00905-w}
}

@book{le2011muon,
  title={Muon spin rotation, relaxation, and resonance: applications to condensed matter},
  author={Le Yaouanc, Alain and De Reotier, Pierre Dalmas},
  number={147},
  year={2011},
  publisher={Oxford Univ. Press}
}

@article{Niggemann_2020,
doi = {10.1088/1361-648X/ab4480},
url = {https://dx.doi.org/10.1088/1361-648X/ab4480},
year = {2019},
month = {oct},
publisher = {IOP Publishing},
volume = {32},
number = {2},
pages = {024001},
author = {Niggemann, Nils and Hering, Max and Reuther, Johannes},
title = {Classical spiral spin liquids as a possible route to quantum spin liquids},
journal = {J. Phys.: Condens. Matter}
}

@article{PhysRevB.109.115110,
  title = {Induced quantum magnetism in crystalline electric field singlet ground state models: Thermodynamics and excitations},
  author = {Thalmeier, Peter and Akbari, Alireza},
  journal = {Phys. Rev. B},
  volume = {109},
  issue = {11},
  pages = {115110},
  numpages = {20},
  year = {2024},
  month = {Mar},
  publisher = {American Physical Society},
  doi = {10.1103/PhysRevB.109.115110},
  url = {https://link.aps.org/doi/10.1103/PhysRevB.109.115110}
}

@article{PhysRevB.96.235113,
  title = {Topological spinon bands and vison excitations in spin-orbit coupled quantum spin liquids},
  author = {Sonnenschein, Jonas and Reuther, Johannes},
  journal = {Phys. Rev. B},
  volume = {96},
  issue = {23},
  pages = {235113},
  numpages = {16},
  year = {2017},
  month = {Dec},
  publisher = {American Physical Society},
  doi = {10.1103/PhysRevB.96.235113},
  url = {https://link.aps.org/doi/10.1103/PhysRevB.96.235113}
}

@article{PhysRevB.58.12049,
  title = {Low-temperature properties of classical geometrically frustrated antiferromagnets},
  author = {Moessner, R. and Chalker, J. T.},
  journal = {Phys. Rev. B},
  volume = {58},
  issue = {18},
  pages = {12049--12062},
  numpages = {0},
  year = {1998},
  month = {Nov},
  publisher = {American Physical Society},
  doi = {10.1103/PhysRevB.58.12049},
  url = {https://link.aps.org/doi/10.1103/PhysRevB.58.12049}
}

@article{PhysRevB.105.094439,
  title = {Timescale distributions of spin fluctuations in the
  $s=2$ kagome antiferromagnet {Cs}{Mn}$_3${F}$_6$({Se}{O$_3$)}$_2$},
  author = {Lee, Suheon and Zhu, Tianyu and Oshima, Y. and Shiroka, T. and Wang, C. and Luetkens, H. and Yang, Haoming and L\"u, Minfeng and Choi, K.-Y.},
  journal = {Phys. Rev. B},
  volume = {105},
  issue = {9},
  pages = {094439},
  numpages = {9},
  year = {2022},
  month = {Mar},
  publisher = {American Physical Society},
  doi = {10.1103/PhysRevB.105.094439},
  url = {https://link.aps.org/doi/10.1103/PhysRevB.105.094439}
}

@Article{Lin2024,
author={Lin, Weijie
and Sheng, Jieming
and Zhao, Nan
and Xiao, Quan
and An, Weiran
and Guo, Ruixin
and Wen, Bo
and Pan, Changzhao
and Wu, Liusuo
and Guo, Shu},
title={Crystal Growth, Structure, and Diverse Magnetic Behaviors in Frustrated Triangular Lattice \ce{REBO3 (RE = Tb--Yb)}},
journal={Inorg. Chem.},
year={2024},
month={Sep},
day={09},
publisher={American Chemical Society},
volume={63},
number={36},
pages={16667-16675},
issn={0020-1669},
doi={10.1021/acs.inorgchem.4c01918},
url={https://doi.org/10.1021/acs.inorgchem.4c01918}
}

@article{PhysRevLett.73.3306,
  title = {Spin Fluctuations in Frustrated Kagom\'e Lattice System \text{Sr}\text{Cr$_{8}$}\text{Ga$_{4}$}\text{O$_{19}$} Studied by Muon Spin Relaxation},
  author = {Uemura, Y. J. and Keren, A. and Kojima, K. and Le, L. P. and Luke, G. M. and Wu, W. D. and Ajiro, Y. and Asano, T. and Kuriyama, Y. and Mekata, M. and Kikuchi, H. and Kakurai, K.},
  journal = {Phys. Rev. Lett.},
  volume = {73},
  issue = {24},
  pages = {3306--3309},
  numpages = {0},
  year = {1994},
  month = {Dec},
  publisher = {American Physical Society},
  doi = {10.1103/PhysRevLett.73.3306},
  url = {https://link.aps.org/doi/10.1103/PhysRevLett.73.3306}
}

@article{PhysRevLett.93.167204,
  title = {Dipolar Spin Correlations in Classical Pyrochlore Magnets},
  author = {Isakov, S. V. and Gregor, K. and Moessner, R. and Sondhi, S. L.},
  journal = {Phys. Rev. Lett.},
  volume = {93},
  issue = {16},
  pages = {167204},
  numpages = {4},
  year = {2004},
  month = {Oct},
  publisher = {American Physical Society},
  doi = {10.1103/PhysRevLett.93.167204},
  url = {https://link.aps.org/doi/10.1103/PhysRevLett.93.167204}
}

@article{PhysRev.79.357,
  title = {Antiferromagnetism. {T}he triangular {I}sing net},
  author = {Wannier, G. H.},
  journal = {Phys. Rev.},
  volume = {79},
  issue = {2},
  pages = {357--364},
  numpages = {0},
  year = {1950},
  month = {Jul},
  publisher = {American Physical Society},
  doi = {10.1103/PhysRev.79.357},
  url = {https://link.aps.org/doi/10.1103/PhysRev.79.357}
}

@Article{Vinograd2022,
author={Vinograd, I.
and Shirer, K. R.
and Massat, P.
and Wang, Z.
and Kissikov, T.
and Garcia, D.
and Bachmann, M. D.
and Horvati{\'{c}}, M.
and Fisher, I. R.
and Curro, N. J.},
title={Second order Zeeman interaction and ferroquadrupolar order in {TmVO}$_4$},
journal={npj Quantum Mater.},
year={2022},
month={Jun},
day={27},
volume={7},
issue={1},
pages={68},
issn={2397-4648},
doi={10.1038/s41535-022-00475-1},
url={https://doi.org/10.1038/s41535-022-00475-1}
}

@article{
doi:10.1073/pnas.2119942119,
author = {Pierre Massat  and Jiajia Wen  and Jack M. Jiang  and Alexander T. Hristov  and Yaohua Liu  and Rebecca W. Smaha  and Robert S. Feigelson  and Young S. Lee  and Rafael M. Fernandes  and Ian R. Fisher },
title = {Field-tuned ferroquadrupolar quantum phase transition in the insulator \text{Tm}\text{V}\text{O$_{4}$}},
journal = {Proc. Natl. Acad. Sci.},
volume = {119},
number = {28},
pages = {e2119942119},
year = {2022},
doi = {10.1073/pnas.2119942119},
URL = {https://www.pnas.org/doi/abs/10.1073/pnas.2119942119}}

@article{PhysRevLett.84.2957,
  title = {Entropy Balance and Evidence for Local Spin Singlets in a Kagom\'e-Like Magnet},
  author = {Ramirez, A. P. and Hessen, B. and Winklemann, M.},
  journal = {Phys. Rev. Lett.},
  volume = {84},
  issue = {13},
  pages = {2957--2960},
  numpages = {0},
  year = {2000},
  month = {Mar},
  publisher = {American Physical Society},
  doi = {10.1103/PhysRevLett.84.2957},
  url = {https://link.aps.org/doi/10.1103/PhysRevLett.84.2957}
}

@article{PhysRevB.110.L020402,
  title = {Classification of classical spin liquids: Typology and resulting landscape},
  author = {Yan, Han and Benton, Owen and Moessner, Roderich and Nevidomskyy, Andriy H.},
  journal = {Phys. Rev. B},
  volume = {110},
  issue = {2},
  pages = {L020402},
  numpages = {6},
  year = {2024},
  month = {Jul},
  publisher = {American Physical Society},
  doi = {10.1103/PhysRevB.110.L020402},
  url = {https://link.aps.org/doi/10.1103/PhysRevB.110.L020402}
}

@article{doi:10.1146/annurev-conmatphys-020911-125138,
  author =        {Witczak-Krempa, William and Chen, Gang and
                   Kim, Yong Baek and Balents, Leon},
  journal =       {Annu. Rev. Condens. Matter Phys.},
  number =        {1},
  pages =         {57-82},
  title =         {Correlated Quantum Phenomena in the Strong Spin-Orbit
                   Regime},
  volume =        {5},
  year =          {2014},
  doi =           {10.1146/annurev-conmatphys-020911-125138},
  url =           {https://doi.org/10.1146/annurev-conmatphys-020911-125138},
}

@article{Balents2010,
  author =        {Balents, Leon},
  journal =       {Nature},
  month =         {Mar},
  number =        {7286},
  pages =         {199-208},
  title =         {Spin liquids in frustrated magnets},
  volume =        {464},
  year =          {2010},
  doi =           {10.1038/nature08917},
  issn =          {1476-4687},
  url =           {https://doi.org/10.1038/nature08917},
}

@article{Jeon2024,
  author =        {Jeon, Sungmin and Wulferding, Dirk and Choi, Youngsu and
                   Lee, Seungyeol and Nam, Kiwan and Kim, Kee Hoon and
                   Lee, Minseong and Jang, Tae-Hwan and Park, Jae-Hoon and
                   Lee, Suheon and Choi, Sungkyun and Lee, Chanhyeon and
                   Nojiri, Hiroyuki and Choi, Kwang-Yong},
  journal =       {Nat. Phys.},
  month =         {Mar},
  number =        {3},
  pages =         {435-441},
  title =         {One-ninth magnetization plateau stabilized by spin
                   entanglement in a kagome antiferromagnet},
  volume =        {20},
  year =          {2024},
  doi =           {10.1038/s41567-023-02318-7},
  issn =          {1745-2481},
  url =           {https://doi.org/10.1038/s41567-023-02318-7},
}

@article{doi:10.1126/science.aay0668,
  author =        {C. Broholm and R. J. Cava and S. A. Kivelson and
                   D. G. Nocera and M. R. Norman and T. Senthil},
  journal =       {Science},
  number =        {6475},
  pages =         {eaay0668},
  title =         {Quantum spin liquids},
  volume =        {367},
  year =          {2020},
  doi =           {10.1126/science.aay0668},
  url =           {https://www.science.org/doi/abs/10.1126/science.aay0668},
}

@article{Khuntia2020,
  author =        {Khuntia, P. and Velazquez, M. and Barth{\'e}lemy, Q. and
                   Bert, F. and Kermarrec, E. and Legros, A. and
                   Bernu, B. and Messio, L. and Zorko, A. and
                   Mendels, P.},
  journal =       {Nat. Phys.},
  month =         {Apr},
  number =        {4},
  pages =         {469-474},
  title =         {Gapless ground state in the archetypal quantum kagome
                   antiferromagnet
                   {Zn\text{Cu$_{3}$}\text{(OH$_{6}$}\text{Cl$_{2}$}}},
  volume =        {16},
  year =          {2020},
  doi =           {10.1038/s41567-020-0792-1},
  issn =          {1745-2481},
  url =           {https://doi.org/10.1038/s41567-020-0792-1},
}

@techreport{khuntia25,
author = {P. Khuntia and D. T. Adroja},
title = {Crystal field excitations in the spin liquid candidate on a stuffed honeycomb lattice antiferromagnet: {TbBO}$_3$},
institution = {STFC ISIS Neutron and Muon Source},
year = {2025},
doi = {https://doi.org/10.5286/ISIS.E.RB2420326},
}

@article{KHATUA20231,
  author =        {J. Khatua and B. Sana and A. Zorko and M. Gomilšek and K. Sethupathi and M. S. Ramachandra Rao and M. Baenitz and B. Schmidt and P. Khuntia},
  journal =       {Phys. Rep.},
  pages =         {1--60},
  title =         {Experimental signatures of quantum and topological
                   states in frustrated magnetism},
  volume =        {1041},
  year =          {2023},
  doi =           {https://doi.org/10.1016/j.physrep.2023.09.008},
  issn =          {0370-1573},
  url =           {https://www.sciencedirect.com/science/article/pii/
                  S037015732300306X},
}

@article{Savary_2016,
  author =        {Lucile Savary and Leon Balents},
  journal =       {Rep. Prog. Phys.},
  month =         {nov},
  number =        {1},
  pages =         {016502},
  publisher =     {{IOP} Publishing},
  title =         {Quantum spin liquids: a review},
  volume =        {80},
  year =          {2016},
  doi =           {10.1088/0034-4885/80/1/016502},
  url =           {https://doi.org/10.1088/0034-4885/80/1/016502},
}

@article{PhysRevLett.69.2590,
  author =        {Bernu, B. and Lhuillier, C. and Pierre, L.},
  journal =       {Phys. Rev. Lett.},
  month =         {Oct},
  pages =         {2590--2593},
  publisher =     {American Physical Society},
  title =         {Signature of {N}\'eel order in exact spectra of quantum
                   antiferromagnets on finite lattices},
  volume =        {69},
  year =          {1992},
  doi =           {10.1103/PhysRevLett.69.2590},
  url =           {https://link.aps.org/doi/10.1103/PhysRevLett.69.2590},
}

@article{PhysRevB.93.144411,
  author =        {Iqbal, Yasir and Hu, Wen-Jun and Thomale, Ronny and
                   Poilblanc, Didier and Becca, Federico},
  journal =       {Phys. Rev. B},
  month =         {Apr},
  pages =         {144411},
  publisher =     {American Physical Society},
  title =         {Spin liquid nature in the {H}eisenberg
                   ${J}_{1}\ensuremath{-}{J}_{2}$ triangular
                   antiferromagnet},
  volume =        {93},
  year =          {2016},
  doi =           {10.1103/PhysRevB.93.144411},
  url =           {https://link.aps.org/doi/10.1103/PhysRevB.93.144411},
}

@article{PhysRevB.92.041105,
  author =        {Zhu, Zhenyue and White, Steven R.},
  journal =       {Phys. Rev. B},
  month =         {Jul},
  pages =         {041105},
  publisher =     {American Physical Society},
  title =         {Spin liquid phase of the
  $S=\frac{1}{2}\phantom{\rule{4.pt}{0ex}}{J}_{1}\ensuremath{-}{J}_{2}$
  {H}eisenberg model on the triangular lattice},
  volume =        {92},
  year =          {2015},
  doi =           {10.1103/PhysRevB.92.041105},
  url =           {https://link.aps.org/doi/10.1103/PhysRevB.92.041105},
}

@article{PhysRevB.94.121111,
  author =        {Saadatmand, S. N. and McCulloch, I. P.},
  journal =       {Phys. Rev. B},
  month =         {Sep},
  pages =         {121111},
  publisher =     {American Physical Society},
  title =         {Symmetry fractionalization in the topological phase
                   of the spin-$\frac{1}{2}$
                   ${J}_{1}\text{\ensuremath{-}}{J}_{2}$ triangular
                   {H}eisenberg model},
  volume =        {94},
  year =          {2016},
  doi =           {10.1103/PhysRevB.94.121111},
  url =           {https://link.aps.org/doi/10.1103/PhysRevB.94.121111},
}

@article{PhysRevB.98.220409,
  author =        {Baenitz, M. and Schlender, Ph. and Sichelschmidt, J. and
                   Onykiienko, Y. A. and Zangeneh, Z. and Ranjith, K. M. and
                   Sarkar, R. and Hozoi, L. and Walker, H. C. and
                   Orain, J.-C. and Yasuoka, H. and van den Brink, J. and
                   Klauss, H. H. and Inosov, D. S. and Doert, Th.},
  journal =       {Phys. Rev. B},
  month =         {Dec},
  pages =         {220409},
  publisher =     {American Physical Society},
  title =         {{NaYb}\text{S$_{2}$}: A planar spin-$\frac{1}{2}$
                   triangular-lattice magnet and putative spin liquid},
  volume =        {98},
  year =          {2018},
  doi =           {10.1103/PhysRevB.98.220409},
  url =           {https://link.aps.org/doi/10.1103/PhysRevB.98.220409},
}

@article{Bordelon2019,
  author =        {Bordelon, Mitchell M. and Kenney, Eric and
                   Liu, Chunxiao and Hogan, Tom and Posthuma, Lorenzo and
                   Kavand, Marzieh and Lyu, Yuanqi and Sherwin, Mark and
                   Butch, N. P. and Brown, Craig and Graf, M. J. and
                   Balents, Leon and Wilson, Stephen D.},
  journal =       {Nat. Phys.},
  month =         {Oct},
  number =        {10},
  pages =         {1058-1064},
  title =         {Field-tunable quantum disordered ground state in the
                   triangular-lattice antiferromagnet {NaYbO$_{2}$}},
  volume =        {15},
  year =          {2019},
  doi =           {10.1038/s41567-019-0594-5},
  issn =          {1745-2481},
  url =           {https://doi.org/10.1038/s41567-019-0594-5},
}

@article{PhysRevX.11.021044,
  author =        {Dai, Peng-Ling and Zhang, Gaoning and Xie, Yaofeng and
                   Duan, Chunruo and Gao, Yonghao and Zhu, Zihao and
                   Feng, Erxi and Tao, Zhen and Huang, Chien-Lung and
                   Cao, Huibo and Podlesnyak, Andrey and
                   Granroth, Garrett E. and Everett, Michelle S. and
                   Neuefeind, Joerg C. and Voneshen, David and
                   Wang, Shun and Tan, Guotai and Morosan, Emilia and
                   Wang, Xia and Lin, Hai-Qing and Shu, Lei and
                   Chen, Gang and Guo, Yanfeng and Lu, Xingye and
                   Dai, Pengcheng},
  journal =       {Phys. Rev. X},
  month =         {May},
  pages =         {021044},
  publisher =     {American Physical Society},
  title =         {Spinon Fermi Surface Spin Liquid in a Triangular
                   Lattice Antiferromagnet {Na}{Yb}\text{Se$_{2}$}},
  volume =        {11},
  year =          {2021},
  doi =           {10.1103/PhysRevX.11.021044},
  url =           {https://link.aps.org/doi/10.1103/PhysRevX.11.021044},
}

@article{KITAEV20062,
  author =        {Alexei Kitaev},
  journal =       {Ann. Phys.},
  number =        {1},
  pages =         {2 - 111},
  title =         {Anyons in an exactly solved model and beyond},
  volume =        {321},
  year =          {2006},
  doi =           {https://doi.org/10.1016/j.aop.2005.10.005},
  issn =          {0003-4916},
  url =           {http://www.sciencedirect.com/science/article/pii/
                  S0003491605002381},
}

@article{Takagi2019,
  author =        {Takagi, Hidenori and Takayama, Tomohiro and
                   Jackeli, George and Khaliullin, Giniyat and
                   Nagler, Stephen E.},
  journal =       {Nat. Rev. Phys.},
  number =        {4},
  pages =         {264-280},
  title =         {Concept and realization of {K}itaev quantum spin
                   liquids},
  volume =        {1},
  year =          {2019},
  doi =           {10.1038/s42254-019-0038-2},
  issn =          {2522-5820},
  url =           {https://doi.org/10.1038/s42254-019-0038-2},
}

@article{PhysRevX.9.031005,
  author =        {Kim, Jaewook and Wang, Xueyun and Huang, Fei-Ting and
                   Wang, Yazhong and Fang, Xiaochen and Luo, Xuan and
                   Li, Y. and Wu, Meixia and Mori, S. and Kwok, D. and
                   Mun, Eun Deok and Zapf, V. S. and Cheong, Sang-Wook},
  journal =       {Phys. Rev. X},
  month =         {Jul},
  pages =         {031005},
  publisher =     {American Physical Society},
  title =         {Spin Liquid State and Topological Structural Defects
                   in Hexagonal {Tb}{In}\text{O$_{3}$}},
  volume =        {9},
  year =          {2019},
  doi =           {10.1103/PhysRevX.9.031005},
  url =           {https://link.aps.org/doi/10.1103/PhysRevX.9.031005},
}

@article{PhysRevB.98.045119,
  title = {Selective measurements of intertwined multipolar orders: {N}on-{K}ramers doublets on a triangular lattice},
  author = {Liu, Changle and Li, Yao-Dong and Chen, Gang},
  journal = {Phys. Rev. B},
  volume = {98},
  issue = {4},
  pages = {045119},
  numpages = {12},
  year = {2018},
  month = {Jul},
  publisher = {American Physical Society},
  doi = {10.1103/PhysRevB.98.045119},
  url = {https://link.aps.org/doi/10.1103/PhysRevB.98.045119}
}

@article{PhysRevB.108.064432,
  title = {Diffuse spin waves and classical spin liquid behavior in the kagom\'e antiferromagnet chromium jarosite, \text{K}\text{Cr$_{3}$}\text{(OD)$_{6}$}\text{(SO$_{4}$)$_{2}$}},
  author = {Holm-Janas, Sofie and Lolk, Sidse L. and Andersen, Anders B. A. and Guidi, Tatiana and Voneshen, David and \ifmmode \check{Z}\else \v{Z}\fi{}ivkovi\ifmmode \acute{c}\else \'{c}\fi{}, Ivica and Nielsen, Ulla Gro and Lefmann, Kim},
  journal = {Phys. Rev. B},
  volume = {108},
  issue = {6},
  pages = {064432},
  numpages = {14},
  year = {2023},
  month = {Aug},
  publisher = {American Physical Society},
  doi = {10.1103/PhysRevB.108.064432},
  url = {https://link.aps.org/doi/10.1103/PhysRevB.108.064432}
}

@article{doi:10.7566/JPSJ.83.034709,
author = {Hattori ,Kazumasa and Tsunetsugu ,Hirokazu},
title = {Antiferro Quadrupole Orders in Non-{K}ramers Doublet Systems},
journal = {J. Phys. Soc. Jpn.},
volume = {83},
number = {3},
pages = {034709},
year = {2014},
doi = {10.7566/JPSJ.83.034709},

URL = { 
    
        https://doi.org/10.7566/JPSJ.83.034709
    
    

}
}

@article{PhysRevResearch.6.023267,
  title = {Excitation spectrum and spin Hamiltonian of the frustrated quantum Ising magnet \text{Pr$_{3}$}\text{B}\text{W}\text{O$_{9}$}},
  author = {Nagl, J. and Flavi\'an, D. and Hayashida, S. and Povarov, K. Yu. and Yan, M. and Murai, N. and Ohira-Kawamura, S. and Simutis, G. and Hicken, T. J. and Luetkens, H. and Baines, C. and Hauspurg, A. and Schwarze, B. V. and Husstedt, F. and Pomjakushin, V. and Fennell, T. and Yan, Z. and Gvasaliya, S. and Zheludev, A.},
  journal = {Phys. Rev. Res.},
  volume = {6},
  issue = {2},
  pages = {023267},
  numpages = {18},
  year = {2024},
  month = {Jun},
  publisher = {American Physical Society},
  doi = {10.1103/PhysRevResearch.6.023267},
  url = {https://link.aps.org/doi/10.1103/PhysRevResearch.6.023267}
}

@Article{Goremychkin2008,
author={Goremychkin, E. A.
and Osborn, R.
and Rainford, B. D.
and Macaluso, R. T.
and Adroja, D. T.
and Koza, M.},
title={Spin-glass order induced by dynamic frustration},
journal={ Nat. Phys.},
year={2008},
month={Oct},
day={01},
volume={4},
number={10},
pages={766-770},
issn={1745-2481},
doi={10.1038/nphys1028},
url={https://doi.org/10.1038/nphys1028}
}

@Article{Tang2023,
author={Tang, Nan
and Gritsenko, Yulia
and Kimura, Kenta
and Bhattacharjee, Subhro
and Sakai, Akito
and Fu, Mingxuan
and Takeda, Hikaru
and Man, Huiyuan
and Sugawara, Kento
and Matsumoto, Yosuke
and Shimura, Yasuyuki
and Wen, Jiajia
and Broholm, Collin
and Sawa, Hiroshi
and Takigawa, Masashi
and Sakakibara, Toshiro
and Zherlitsyn, Sergei
and Wosnitza, Joachim
and Moessner, Roderich
and Nakatsuji, Satoru},
title={Spin-orbital liquid state and liquid-gas metamagnetic transition on a pyrochlore lattice},
journal={Nat. Phys.},
year={2023},
month={Jan},
day={01},
volume={19},
issue={1},
pages={92--98},
issn={1745-2481},
doi={10.1038/s41567-022-01816-4},
url={https://doi.org/10.1038/s41567-022-01816-4}
}

@article{Clark2019,
  author =        {Clark, Lucy and Sala, Gabriele and Maharaj, Dalini D. and
                   Stone, Matthew B. and Knight, Kevin S. and
                   Telling, Mark T. F. and Wang, Xueyun and Xu, Xianghan and
                   Kim, Jaewook and Li, Yanbin and Cheong, Sang-Wook and
                   Gaulin, Bruce D.},
  journal =       {Nat. Phys.},
  number =        {3},
  pages =         {262-268},
  title =         {Two-dimensional spin liquid behaviour in the
                   triangular-honeycomb antiferromagnet
                   {Tb}{In}\text{O$_{3}$}},
  volume =        {15},
  year =          {2019},
  doi =           {10.1038/s41567-018-0407-2},
  issn =          {1745-2481},
  url =           {https://doi.org/10.1038/s41567-018-0407-2},
}

@misc{sm,
  note =          {See supplementary material at  [\text{URL} to be provided by the publisher]  for further
                   details on sample characterization, crystal electric field,  additional experimental results, and DFT calculations,  along with Refs. [2-5, 9, 29].},
  OPTyear = {2025},
}

@article{Lin2004,
  author =        {Lin, Jianhua and Sheptyakov, Denis and Wang, Yingxia and
                   Allenspach, Peter},
  journal =       {Chem. Mater.},
  month =         {Jun},
  number =        {12},
  pages =         {2418-2424},
  publisher =     {American Chemical Society},
  title =         {Structures and Phase Transition of Vaterite-Type Rare
                   Earth Orthoborates:{\thinspace} A Neutron Diffraction
                   Study},
  volume =        {16},
  year =          {2004},
  doi =           {10.1021/cm0499388},
  issn =          {0897-4756},
  url =           {https://doi.org/10.1021/cm0499388},
}

@article{MUKHERJEE2018173,
  author =        {P. Mukherjee and Y. Wu and G.I. Lampronti and
                   S.E. Dutton},
  journal =       {Mater. Res. Bull.},
  pages =         {173-179},
  title =         {Magnetic properties of monoclinic lanthanide
                   orthoborates, {Ln}{B}\text{O$_{3}$}, {Ln}= {Gd},
                   {Tb}, {Dy}, {Ho}, {Er}, {Yb}},
  volume =        {98},
  year =          {2018},
  doi =           {https://doi.org/10.1016/j.materresbull.2017.10.007},
  issn =          {0025-5408},
  url =           {https://www.sciencedirect.com/science/article/pii/
                  S0025540817324558},
}

@article{Sarte2021,
  author =        {Sarte, P. M. and Cruz-Kan, K. and Ortiz, B. R. and
                   Hong, K. H. and Bordelon, M. M. and
                   Reig-i-Plessis, D. and Lee, M. and Choi, E. S. and
                   Stone, M. B. and Calder, S. and Pajerowski, D. M. and
                   Mangin-Thro, L. and Qiu, Y. and Attfield, J. P. and
                   Wilson, S. D. and Stock, C. and Zhou, H. D. and
                   Hallas, A. M. and Paddison, J. A. M. and Aczel, A. A. and
                   Wiebe, C. R.},
  journal =       {npj Quantum Mater.},
  month =         {May},
  number =        {1},
  pages =         {42},
  title =         {Dynamical ground state in the {XY} pyrochlore
                   {Yb$_{2}$}{GaSb}{O$_{7}$}},
  volume =        {6},
  year =          {2021},
  doi =           {10.1038/s41535-021-00343-4},
  issn =          {2397-4648},
  url =           {https://doi.org/10.1038/s41535-021-00343-4},
}

@phdthesis{mukherjee_2018,
  author =        {Mukherjee, Paromita},
  publisher =     {Apollo - University of Cambridge Repository},
  title =         {Investigation of the magnetic and magnetocaloric
                   properties of complex lanthanide oxides},
  year =          {2018},
  month =         {May},
  school =        {Cambridge University},
  doi =           {10.17863/CAM.22636},
  url =           {https://www.repository.cam.ac.uk/handle/1810/275425},
}

@article{PhysRevB.100.144432,
  author =        {Ding, Lei and Manuel, Pascal and Bachus, Sebastian and
                   Gru\ss{}ler, Franziska and Gegenwart, Philipp and
                   Singleton, John and Johnson, Roger D. and
                   Walker, Helen C. and Adroja, Devashibhai T. and
                   Hillier, Adrian D. and Tsirlin, Alexander A.},
  journal =       {Phys. Rev. B},
  month =         {Oct},
  pages =         {144432},
  publisher =     {American Physical Society},
  title =         {Gapless spin-liquid state in the structurally
                   disorder-free triangular antiferromagnet
                   {Na}{Yb}\text{O$_{2}$}},
  volume =        {100},
  year =          {2019},
  doi =           {10.1103/PhysRevB.100.144432},
  url =           {https://link.aps.org/doi/10.1103/PhysRevB.100.144432},
}

@article{PhysRevLett.116.107203,
  author =        {Khuntia, P. and Bert, F. and Mendels, P. and
                   Koteswararao, B. and Mahajan, A. V. and Baenitz, M. and Chou, F. C. and Baines, C. and Amato, A. and Furukawa, Y.},
  journal =       {Phys. Rev. Lett.},
  month =         {Mar},
  pages =         {107203},
  publisher =     {American Physical Society},
  title =         {Spin Liquid State in the {3D} Frustrated
                   Antiferromagnet
                   {Pb}{Cu}{Te}$_2${O}$_6$: {NMR} and
                   Muon Spin Relaxation Studies},
  volume =        {116},
  year =          {2016},
  doi =           {10.1103/PhysRevLett.116.107203},
  url =          {https://link.aps.org/doi/10.1103/PhysRevLett.116.107203},
}

@article{PhysRevLett.109.037208,
  author =        {F\aa{}k, B. and Kermarrec, E. and Messio, L. and
                   Bernu, B. and Lhuillier, C. and Bert, F. and
                   Mendels, P. and Koteswararao, B. and Bouquet, F. and
                   Ollivier, J. and Hillier, A. D. and Amato, A. and
                   Colman, R. H. and Wills, A. S.},
  journal =       {Phys. Rev. Lett.},
  month =         {Jul},
  pages =         {037208},
  publisher =     {American Physical Society},
  title =         {Kapellasite: A Kagome Quantum Spin Liquid with
                   Competing Interactions},
  volume =        {109},
  year =          {2012},
  doi =           {10.1103/PhysRevLett.109.037208},
  url =           {https://link.aps.org/doi/10.1103/PhysRevLett.109.037208},
}

@article{PhysRevB.75.094404,
  author =        {Baker, P. J. and Blundell, S. J. and Pratt, F. L. and
                   Lancaster, T. and Brooks, M. L. and Hayes, W. and
                   Isobe, M. and Ueda, Y. and Hoinkis, M. and Sing, M. and
                   Klemm, M. and Horn, S. and Claessen, R.},
  journal =       {Phys. Rev. B},
  month =         {Mar},
  pages =         {094404},
  publisher =     {American Physical Society},
  title =         {Muon spin relaxation measurements on the dimerized
                   spin 1/2 chains {NaTi}\text{Si$_{2}$}\text{O$_{6}$}
                   {and} {TiOCl}},
  volume =        {75},
  year =          {2007},
  doi =           {10.1103/PhysRevB.75.094404},
  url =           {https://link.aps.org/doi/10.1103/PhysRevB.75.094404},
}

@article{doi:10.1143/JPSJ.55.1751,
  author =        {Yamada ,Yoshihiro and Sakata ,Akihiko},
  journal =       {J. Phys. Soc. Jpn.},
  number =        {5},
  pages =         {1751-1758},
  title =         {An Analysis Method of Antiferromagnetic Powder
                   Patterns in Spin-Echo {NMR} under External Fields},
  volume =        {55},
  year =          {1986},
  doi =           {10.1143/JPSJ.55.1751},
  url =           {https://doi.org/10.1143/JPSJ.55.1751},
}

@article{PhysRevLett.104.057202,
  title = {Ground State of the Easy-Axis Rare-Earth Kagome Langasite \ce{Pr3Ga5SiO14}},
  author = {Zorko, A. and Bert, F. and Mendels, P. and Marty, K. and Bordet, P.},
  journal = {Phys. Rev. Lett.},
  volume = {104},
  issue = {5},
  pages = {057202},
  numpages = {4},
  year = {2010},
  month = {Feb},
  publisher = {American Physical Society},
  doi = {10.1103/PhysRevLett.104.057202},
}

@article{PhysRevB.102.045149,
  author =        {Zeng, K. Y. and Ma, Long and Gao, Y. X. and
                   Tian, Z. M. and Ling, L. S. and Pi, Li},
  journal =       {Phys. Rev. B},
  month =         {Jul},
  pages =         {045149},
  publisher =     {American Physical Society},
  title =         {NMR study of the spin excitations in the frustrated
                   antiferromagnet
                   $\mathrm{Yb}(\mathrm{BaBO}{}_{3}{)}_{3}$ with a
                   triangular lattice},
  volume =        {102},
  year =          {2020},
  doi =           {10.1103/PhysRevB.102.045149},
  url =           {https://link.aps.org/doi/10.1103/PhysRevB.102.045149},
}

\end{document}